%% file: beamE_draft.tex
\documentclass[cpc,amsmath,amssymb,showpacs,superscriptaddress,nofootinbib,twocolumn]{revtex4-1}
\usepackage{graphicx}
\usepackage{epsfig,graphics,subfigure,psfrag,amsmath,amssymb}
\usepackage[mathlines]{lineno}
\usepackage{dcolumn}
\usepackage{color}
\usepackage{bm}
\usepackage{CJK,CJKnumb,CJKulem}
\usepackage{overpic}
\usepackage{rotating}
\usepackage{times}
\usepackage{multirow}
\usepackage{float}
\usepackage{mathrsfs}
\usepackage{booktabs}
\usepackage[mathlines]{lineno}
\newcommand{\tabincell}[2]{\begin{tabular}{@{}#1@{}}#2\end{tabular}}
\begin{document}

\begin{CJK*}{GBK}{song}
\title{\boldmath Measurements of the center-of-mass energies at BESIII via the di-muon process}

\input{author_20150731}

\begin{abstract}
  From 2011 to 2014, the BESIII experiment collected about 5 fb$^{-1}$
  data at center-of-mass energies around 4 GeV for the studies of the
  charmonium-like and higher excited charmonium states. By analyzing
  the di-muon process $e^{+}e^{-}\rightarrow\gamma_{\rm
    ISR/FSR}\mu^{+}\mu^{-}$, the center-of-mass energies of the data
  samples are measured with a precision of 0.8 MeV. The center-of-mass
  energy is found to be stable for most of time during the data taking.
\end{abstract}

\maketitle

\section{\boldmath Introduction}

The BESIII detector operating at the BEPCII accelerator is designed to
study physics in the $\tau$-charm energy region (2$\sim$4.6
GeV)~\cite{taucharm}. From 2011 to 2014, the BESIII experiment
accumulated 5 fb$^{-1}$ of $e^+e^-$ collision data at center-of-mass energies between 3.810
and 4.600 GeV to study the charmonium-like and higher excited
charmonium states~\cite{Charmoniumlike}. In the past, BESIII has taken
large data samples at the $J/\psi$, $\psi(3686)$ and $\psi(3770)$
peaks.  The corresponding beam energy was fine tuned by a $J/\psi$ or
$\psi(3686)$ mass scan before the data-taking. However, around 4 GeV,
there is no narrow resonance in $e^{+}e^{-}$ annihilation, and the
$\psi(3686)$ peak is too far away to be used to calibrate the beam
energy. The Beam Energy Measurement System (BEMS), which was installed
in 2008,
is designed to measure the beam energy with a relative systematic
uncertainty of $2\times10^{-5}$~\cite{BEMS} based on the energies of
Compton back-scattered photons. The performance of the BEMS is
verified through the measurement of the $\psi(3686)$ mass, but 4 GeV
is beyond the working range of BEMS. To precisely measure the masses
of the newly observed $Z_c$~\cite{Z3900,Z4020} particles,
especially for those which are observed by a partial reconstruction
method~\cite{Z3885,Z4025}, a precise knowledge of the center-of-mass
energy ($E_{\rm cms}$) is crucial.

In this paper, we develop a method to measure the $E_{\rm cms}$ using
the di-muon process
\begin{equation}\label{Rdimu}
  e^{+}e^{-}\rightarrow\gamma_{\rm ISR/FSR}\mu^{+}\mu^{-},
\end{equation}
\noindent where $\gamma_{\rm ISR/FSR}$ represents possible initial
state radiative (ISR) or final state radiative (FSR) photons. The
$E_{\rm cms}$ can be written as
\begin{equation}\label{E_Rdimu}
  E_{\rm cms} = M(\mu^{+}\mu^{-}) +\Delta M_{\rm ISR/FSR},
\end{equation}
\noindent where $M(\mu^{+}\mu^{-})$ is the invariant mass of
$\mu^{+}\mu^{-}$, $\Delta M_{\rm ISR/FSR}$ is the mass shift due to
ISR/FSR radiation, which equals to the difference between the
invariant mass of the $\mu^{+}\mu^{-}$ pair and the $E_{\rm cms}$ of the
initial $e^{+}e^{-}$ system. In the analysis, $\Delta M_{\rm ISR/FSR}$
is estimated from a Monte Carlo (MC) simulation of the di-muon
process by turning on or off the ISR/FSR, where the ISR/FSR is
simulated by BABAYAGA3.5~\cite{babayaga}. To make sure the measured
invariant mass $M(\mu^{+}\mu^{-})$ is unbiased, we validate the
reconstructed momentum of $\mu^{+}/\mu^{-}$ with the $J/\psi$ signal
from the process $e^{+}e^{-}\rightarrow\gamma_{\rm ISR}J/\psi$ with
$J/\psi\rightarrow\mu^{+}\mu^{-}(\gamma_{\rm FSR})$ in the same data
samples.

\section{\boldmath{The BESIII Detector and Data Sets}}

The BESIII detector is described in detail in Ref.~\cite{BESIII}. The
detector is cylindrically symmetric and covers 93\% of the solid
angle around the collision point. The detector consists of four main
components: (a) A 43-layer main drift chamber (MDC) provides momentum
measurement for charged tracks with a momentum resolution of 0.5\% at 1
GeV/$c$ in a 1 T magnetic field. (b) A time-of-flight system (TOF)
composed of plastic scintillators has a time resolution of 80 ps (110
ps) in the barrel (endcaps). (c) An electromagnetic calorimeter
(EMC) made of 6240 CsI(Tl) crystals provides an energy resolution for
photons of 2.5\% (5\%) at 1 GeV in the barrel (endcaps).  (d) A muon
counter (MUC), consisting of 9 (8) layers of resistive plate chambers
in the barrel (endcaps) within the return yoke of the magnet, provides
2 cm position resolution. The electron and positron beams collide with
an angle of 22 mrad at the interaction point (IP) in order to separate the $e^{+}$
and $e^{-}$ beams after the collision.  A {\sc geant4}~\cite{GEANT4} based detector
simulation package is developed to model the detector response for MC
events.

In total, there are 25 data samples taken at different center-of-mass
energies or during different periods, as listed in
Table~\ref{MDimu_list}. The data sets are listed chronologically, and
the ID number is the requested $E_{\rm cms}$. The offline luminosity
is measured through large-angle Bhabha scattering events with a
precision of 1\%~\cite{Lum_XYZ}. In this paper, we measure $E_{\rm
  cms}$ for all the 25 data samples and examine its stability during
each data taking period.

%\begin{widetext}
\begin{table*}[t]
  \centering
  \caption{\label{MDimu_list}Summary of the data sets, including ID,
    run number, offline luminosity, the measured $M^{\rm
      cor}(J/\psi)$, $M^{\rm obs}(\mu^{+}\mu^{-})$, and $E_{\rm
      cms}$. The first uncertainty is statistical, and the second is
    systematic. Superscripts indicate separate samples acquired at the
    same $E_{\rm cms}$. The "-" indicates samples which are
    combined with the previous one(s) to measure $M^{\rm
      cor}(\mu^{+}\mu^{-})$.}
\begin{tabular}{cccccccc}
\hline
  ID&Run number&Offline lum. (pb$^{-1}$)& $M^{\rm cor}(J/\psi)$ (MeV/$c^{2}$)&&$M^{\rm obs}(\mu^{+}\mu^{-})$ (MeV/$c^{2}$)&& $E_{\rm cms}$ (MeV)\\
\hline
$4009^{1}$&23463 to 23505&\multirow{2}{0.7in}{481.96$\pm$0.01}&3097.00$\pm$0.15  && 4005.90$\pm$0.15 &&4009.10$\pm$0.15$\pm$0.59\\
$4009^{2}$&23510 to 24141&&-&&4004.26$\pm$0.05&&4007.46$\pm$0.05$\pm$0.66\\
$4260^{1}$&29677 to 29805&\multirow{2}{0.7in}{523.74$\pm$0.10}&3096.95$\pm$0.26   &&\tabincell{c}{$(4367.37\!-\!3.75\!\times\!10^{-3}\!\times\! N_{\rm run})$\\$\pm$0.12}   &&\tabincell{c}{$(4370.95\!-\!3.75\!\times\!10^{-3}\!\times\! N_{\rm run})$\\$\pm$0.12$\pm$0.62 }  \\
$4260^{2}$&29822 to 30367&   &  - &&4254.42$\pm$0.06   &&4258.00$\pm$0.06$\pm$0.60   \\
4190     &30372 to 30437     &43.09$\pm$0.03&3097.53$\pm$0.51     &&4185.12$\pm$0.15   &&4188.59$\pm$0.15$\pm$0.68   \\
4230$^{1}$&30438 to 30491    &44.40$\pm$0.03&       -             &&4223.83$\pm$0.18   &&4227.36$\pm$0.18$\pm$0.63   \\
4310     &30492 to 30557     &44.90$\pm$0.03&       -             &&4304.22$\pm$0.17   &&4307.89$\pm$0.17$\pm$0.63  \\
4360     &30616 to 31279     &539.84$\pm$0.10&3096.42$\pm$0.28   &&4354.51$\pm$0.05   &&4358.26$\pm$0.05$\pm$0.62   \\
4390     &31281 to 31325     &55.18$\pm$0.04&3096.39$\pm$0.62    &&4383.60$\pm$0.17   &&4387.40$\pm$0.17$\pm$0.65   \\
4420$^{1}$&31327 to 31390    &44.67$\pm$0.03&    -               &&4413.10$\pm$0.20   &&4416.95$\pm$0.20$\pm$0.63  \\
4260$^{3}$&31561 to 31981    &301.93$\pm$0.08&3096.76$\pm$0.34   &&4253.85$\pm$0.07   &&4257.43$\pm$0.07$\pm$0.66   \\
4210     &31983 to 32045     &54.55$\pm$0.03&3096.88$\pm$0.43    &&4204.23$\pm$0.14   &&4207.73$\pm$0.14$\pm$0.61   \\
4220     &32046 to 32140     &54.13$\pm$0.03&      -              &&4213.61$\pm$0.14   &&4217.13$\pm$0.14$\pm$0.67    \\
4245     &32141 to 32226     &55.59$\pm$0.04&      -              &&4238.10$\pm$0.12   &&4241.66$\pm$0.12$\pm$0.73  \\
$4230^{2}$&32239 to 32849&\multirow{2}{0.7in}{1047.34$\pm$0.14}&3096.58$\pm$0.18&& \tabincell{c}{$(4316.81\!-\!2.87\!\times\!10^{-3}\!\times\! N_{\rm run})$\\$\pm$0.05} && \tabincell{c}{$(4320.34\!-\!2.87\!\times\!10^{-3}\!\times\! N_{\rm run})$\\$\pm$0.05$\pm$0.60}   \\
$4230^{3}$&32850 to 33484&& - &&4222.01$\pm$0.05   &&4225.54$\pm$0.05$\pm$0.65 \\
3810     &33490 to 33556     &50.54$\pm$0.03&3097.38$\pm$0.37    &&3804.82$\pm$0.10   &&3807.65$\pm$0.10$\pm$0.58   \\
3900     &33572 to 33657     &52.61$\pm$0.03&   -                 &&3893.26$\pm$0.11   &&3896.24$\pm$0.11$\pm$0.72   \\
4090     &33659 to 33719     &52.63$\pm$0.03&   -                 &&4082.15$\pm$0.14   &&4085.45$\pm$0.14$\pm$0.66   \\
4600     &35227 to 36213     &566.93$\pm$0.11&3096.54$\pm$0.33   &&4595.38$\pm$0.07   &&4599.53$\pm$0.07$\pm$0.74  \\
4470     &36245 to 36393     &109.94$\pm$0.04&3096.69$\pm$0.42   &&4463.13$\pm$0.11   &&4467.06$\pm$0.11$\pm$0.73  \\
4530     &36398 to 36588     &109.98$\pm$0.04&     -              &&4523.10$\pm$0.11   &&4527.14$\pm$0.11$\pm$0.72   \\
4575     &36603 to 36699     &47.67$\pm$0.03&      -              &&4570.39$\pm$0.18   &&4574.50$\pm$0.18$\pm$0.70   \\
$4420^{2}$&36773 to 37854&\multirow{2}{0.8in}{1028.89$\pm$0.13}&3096.65$\pm$0.21   &&4411.99$\pm$0.04   && 4415.84$\pm$0.04$\pm$0.62      \\
$4420^{3}$&37855 to 38140&& - &&4410.21$\pm$0.07   &&4414.06$\pm$0.07$\pm$0.72  \\
\hline

\end{tabular}
\end{table*}
%\end{widetext}

\section{\boldmath{Muon Momentum validation with $J/\psi$ signal}}\label{Jpsi_cali}

\begin{figure}[!htbp]
  \centering
   \includegraphics[width=0.4\textwidth]{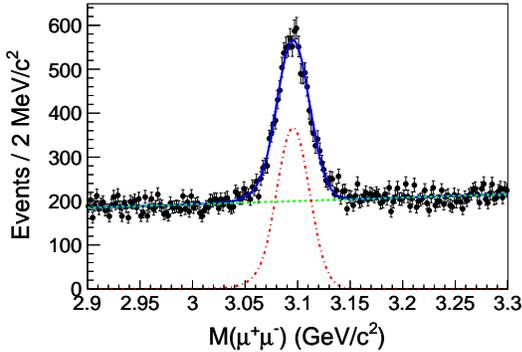}
  \caption{\small{Fit to the $M(\mu^{+}\mu^{-})$ distribution in
      $e^{+}e^{-}\rightarrow\gamma_{\rm ISR}J/\psi$ events for the data sample
      4600. Black dots with error bars are data, the blue curve shows the fit
      result, the red dash-dotted curve is for signal, and the green dashed line is for
      background.}}
  \label{MJpsi_fit} %% label for entire figure
\end{figure}

\begin{figure}[!htbp]
  \centering
   \includegraphics[width=0.4\textwidth]{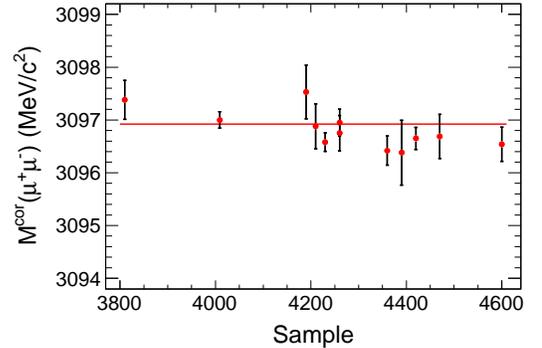}
  \caption{\small{Measured $J/\psi$ mass after the FSR correction,
      $M^{\rm cor}(\mu^{+}\mu^{-})$, for data taken at different
      energies, in which the data samples with small statistics are
      merged (described in text). The red solid line is the nominal $J/\psi$
      mass for reference.}}
  \label{MJpsi_all} %% label for entire figure
\end{figure}

The high momentum measurement of muons is validated with
$J/\psi\rightarrow\mu^{+}\mu^{-}$ candidates selected via the process
$e^{+}e^{-}\rightarrow\gamma_{\rm ISR}J/\psi$. Events must have only
two good oppositely charged tracks. Each good charged track is
required to be
consistent with originating from the IP within $1\rm~cm$ in radial
direction ($V_{xy} < 1$ cm) and $10\rm~cm$ in $z$ direction ($|V_{z}|
< 10$ cm) to the run-dependent IP, and within the polar angle region
$|\cos\theta|<0.8$ (i.e. accepting only tracks in the barrel
region). The energy deposition in the EMC ($E$) for each charged track is
required to be less than 0.4 GeV to suppress background from radiative Bhabha
events. A further requirement on the opening angle between the two
tracks, $\cos\theta_{\mu^{+}\mu^{-}} > -0.98$, is used to remove
cosmic rays. The background remaining after the above selection
comes from the radiative di-muon process, which has exactly the same final state and
can not be completely removed. The radiative di-muon events show a smooth distribution
in $M(\mu^{+}\mu^{-})$.
With the above selection criteria imposed, the distribution of
$M(\mu^{+}\mu^{-})$ of each sample is fitted with a
crystal-ball function~\cite{CB_func} for the $J/\psi$ signal and a
linear function to model the background. Figure~\ref{MJpsi_fit} shows the fit result for the
data sample 4600 as an example. In order to reduce the fluctuation of
$M(\mu^{+}\mu^{-})$, adjacent data samples with small statistics are
combined.  Due to final state radiation,
$J/\psi\rightarrow\mu^{+}\mu^{-}\gamma_{\rm FSR}$, the measured
$M^{\rm obs}(\mu^{+}\mu^{-})$ is slightly lower than the nominal
$J/\psi$ mass~\cite{PDG}. The mass shift due to the FSR photon(s)
$\Delta M_{\rm FSR}$ is estimated by simulated samples of the process
$e^{+}e^{-}\rightarrow\gamma_{\rm ISR}J/\psi$
with 50,000 events each, generated at different energies
using the generator PHOTOS~\cite{Photos} with FSR turned on and
off. The mass shift $\Delta M_{\rm FSR}$ at each $E_{\rm cms}$ is obtained as the difference
in $M^{\rm obs}(\mu^{+}\mu^{-})$ between the MC samples with FSR
turned on and off. These simulation studies validate that $\Delta
M_{\rm FSR}$ is independent of $E_{\rm cms}$. A weighted average, $\overline{\Delta
 M}_{\rm FSR} = (0.59\pm0.04)$ MeV/$c^{2}$, is obtained by fitting
the $\Delta M_{\rm FSR}$ versus $E_{\rm cms}$. The measured mass corrected
by $\overline{\Delta M}_{\rm FSR}$, $M^{\rm cor}(\mu^{+}\mu^{-})$,
is plotted in Fig.~\ref{MJpsi_all} and listed in
Table~\ref{MDimu_list} (column 4). The values of $M^{\rm cor}(\mu^{+}\mu^{-})$
for the different data samples are consistent within errors, and
the average is $\overline{M}^{\rm cor}(\mu^{+}\mu^{-}) =
3096.79\pm0.08$ MeV/$c^{2}$, which agrees with the nominal $J/\psi$
mass within errors. The small difference is considered as systematic
uncertainty in Section~\ref{Sys_err}.

\section{\boldmath The mass shift $\Delta M_{\rm ISR/FSR}$}\label{correct_dM}

The $E_{\rm cms}$ of the initial $e^{+}e^{-}$ pair is measured via the
di-muon process $e^{+}e^{-}\rightarrow\gamma_{\rm
  ISR/FSR}\mu^{+}\mu^{-}$. However, due to the
emission of radiative photons, the invariant mass of the
$\mu^{+}\mu^{-}$ pair is less than the $E_{\rm cms}$ of the initial
$e^{+}e^{-}$ pair by $\Delta M_{\rm ISR/FSR}$. In general, the mass
shift due to the FSR is small, about 0.6 MeV/$c^{2}$ at 3.097 GeV, and
the mass shift due to the ISR is 2$\sim$3 MeV, which has been well
studied theoretically~\cite{QED}. In the analysis, the $\Delta M_{\rm
  ISR/FSR}$ is estimated with MC simulation using
{\sc babayaga3.5}~\cite{babayaga}. We generate 50,000 di-muon MC
  events for each sample with ISR/FSR turned on and off, and take the
difference in $M(\mu^{+}\mu^{-})$ as the mass shift $\Delta M_{\rm
  ISR/FSR}$ caused by ISR and FSR. In order to avoid possible bias,
the same event selection criteria for the di-muon process applied for
data (as described in Section~\ref{Ecms_measure}) are imposed to the MC samples.

\begin{figure}
  \centering
   \includegraphics[width=0.4\textwidth]{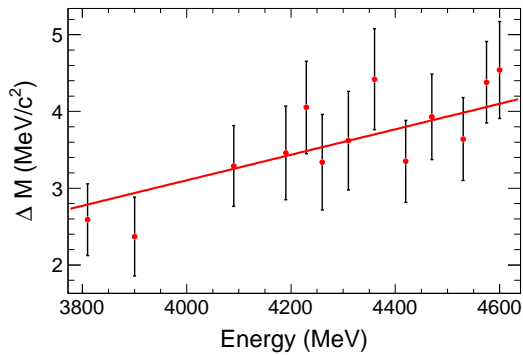}
  \caption{\small{Difference in $M(\mu^{+}\mu^{-})$ between the
      MC samples with ISR/FSR turned on and off (the mass shift $\Delta M_{\rm
        ISR/FSR}$) versus center-of-mass energy for
      $e^+e^-\to\gamma_{\rm ISR/FSR}\mu^+\mu^-$ MC samples. The red solid
      line is the fit result.}}
  \label{dDimu_all_MC} %% label for entire figure
\end{figure}

The distributions of $M(\mu^{+}\mu^{-})$ with ISR/FSR on and off are
fitted with a Gaussian function in a range around the peak (same
method with data in Section~\ref{Ecms_measure}). The difference in
peak positions (the mass shift $\Delta M_{\rm ISR/FSR}$) versus
$E_{\rm cms}$ is seen to increase with $E_{\rm cms}$, as shown in
Fig.~\ref{dDimu_all_MC}.  The $\Delta M_{\rm ISR/FSR}$ is fitted with a
linear function,
$\overline{\Delta M}_{\rm ISR/FSR} =
(-3.53\pm1.11)+(1.67\pm0.28)\times 10^{-3} \times E_{\rm cms}/\text{MeV}$;
the goodness of the fit is $\chi^{2}/n.d.f = 6.3/13$. The resulting
$E_{\rm cms}$-dependent $\overline{\Delta M}_{\rm ISR/FSR}$ will be
used to correct the measured $M^{\rm obs}(\mu^{+}\mu^{-})$ for the
effects of ISR and FSR.

The mass shift due to FSR only, $\Delta M_{\rm FSR}$, is estimated by
comparing MC samples of di-muon production with FSR turned on and off. We find that
$\Delta M_{\rm FSR}$ increases with $E_{\rm cms}$ and we parameterize
the  $E_{\rm cms}$ dependence with a first-order polynomial as
$\overline{\Delta M}_{\rm FSR} = (-1.34\pm0.84)+(0.56\pm0.21)\times
10^{-3}\times E_{\rm cms}$,
where $E_{\rm cms}$ is in unit of MeV and the error matrix of the fit
parameters is
$(0.693, -0.170\times10^{-3},-0.170\times10^{-3},
0.042\times10^{-6})$.
So the corresponding $\Delta M_{\rm FSR}$ at 3.81 GeV (4.6 GeV) is
0.79$\pm$0.09 MeV (1.24$\pm$0.14 MeV).

\section{\boldmath The measurement of $E_{\rm cms}$}\label{Ecms_measure}

To select the di-muon process
$e^{+}e^{-}\rightarrow\gamma_{ISR/FSR}\mu^{+}\mu^{-}$, the requirement
for charged tracks is the same as the $\gamma_{\rm ISR}J/\psi$
selection. To achieve best precision, only events with both tracks in
the barrel region (i.e., in the polar angle region
$|\cos\theta|<0.80$) are accepted. A requirement on the opening angle
between the two tracks of
$178.60^{\circ} < \theta_{\mu\mu} < 179.64^{\circ}$ is applied to
suppress cosmic ray and di-muon events with high-energy radiative
photons. To further remove cosmic ray events, the TOF timing
difference between the two tracks is required to be $|\Delta t| < 4$ ns. The
background contribution following above selection criteria is less than
0.001\% compared to signal and is therefore neglected in the following.

\begin{figure}[!htbp]
  \centering
   \includegraphics[width=0.4\textwidth]{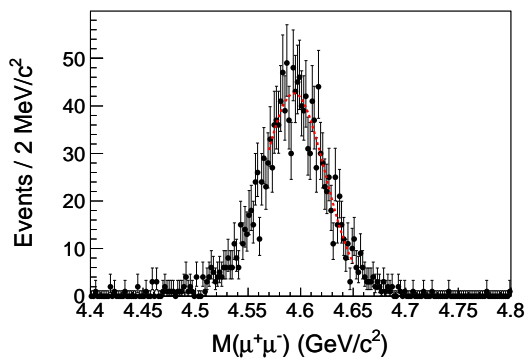}
   \caption{\small{Fit to the $M(\mu^{+}\mu^{-})$ distribution for the data sample
       4600. Black dots with error bars are data, and the red curve shows the
       fit result.}}
  \label{Mdata_fit} %% label for entire figure
\end{figure}

\begin{figure*}[!htbp]
  \centering
   \includegraphics[width=0.99\textwidth]{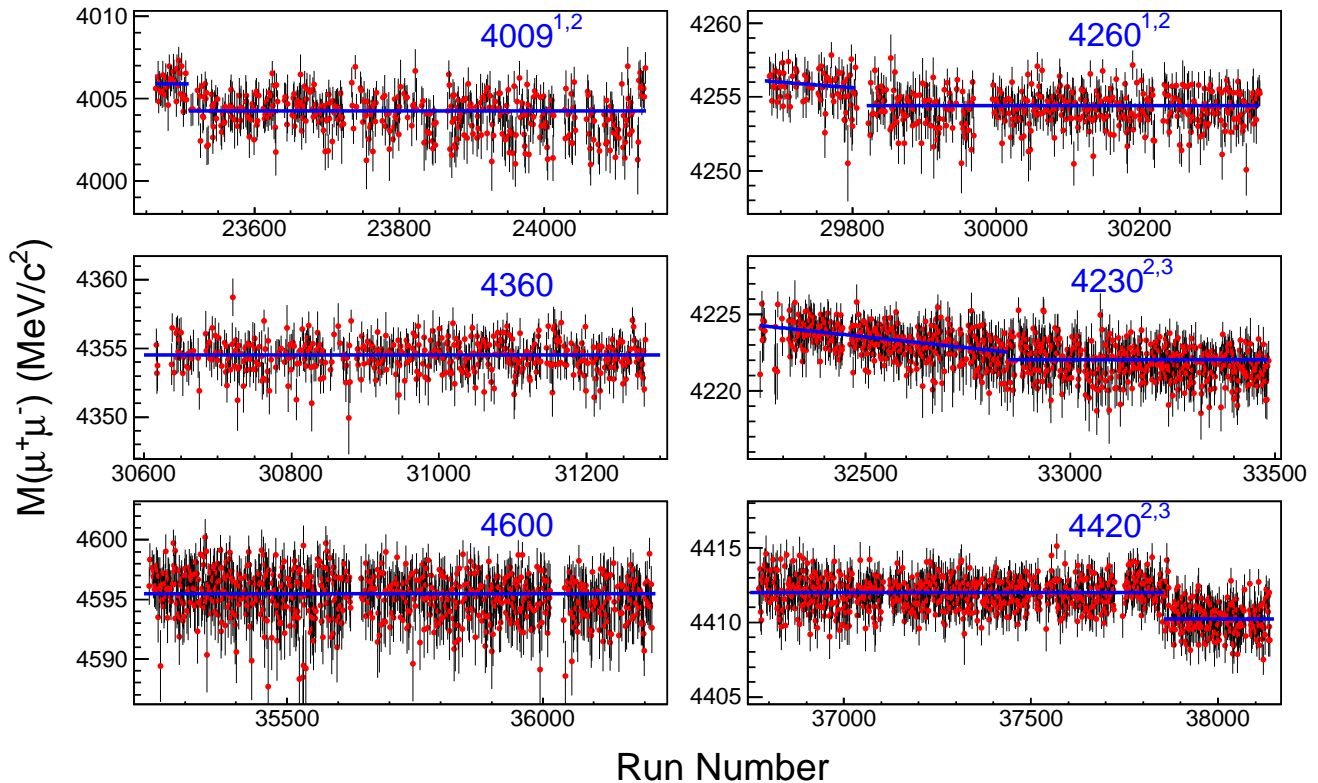}
   \caption{\small{Measured $M(\mu^{+}\mu^{-})$ of di-muon events
       run-by-run for samples 4009$^{1,2}$, 4260$^{1,2}$, 4360,
       4230$^{2,3}$, 4600, and 4420$^{2,3}$. The blue solid lines show the
       fit results for the data samples.}}
  \label{run_by_run} %% label for entire figure
\end{figure*}

We estimate the peak position of the distribution of
$M^{\rm obs}(\mu^{+}\mu^{-})$ for selected di-muon events by fitting
with a Gaussian function in the range of $(-1\sigma,2\sigma)$ around
the peak, where $\sigma$ is the standard deviation of the Gaussian. To
examine the stability of the $E_{\rm cms}$ over time for each data
sample, the fit procedure is performed for each run of the data
samples, where a run normally corresponds to one hour of data taking. The fit
result for one run of the 4600 data sample is shown in Fig.~\ref{Mdata_fit}. The
measured $\mu^{+}\mu^{-}$ masses versus the run number for the samples
4009$^{1,2}$, 4260$^{1,2}$, 4360, 4230$^{2,3}$, 4600, and 4420$^{2,3}$
are plotted in Fig.~\ref{run_by_run}.  For the sample 4260$^{1}$
(4230$^{2}$), the measured $M^{\rm obs}(\mu^{+}\mu^{-})$ changes
slowly and is fitted with a linear function. The fit gives
$(4367.37\pm53.53)+(-3.75\pm1.80)\times10^{-3}\times N_{\rm run}$
($(4316.81\pm7.76)+(-2.87\pm0.25)\times10^{-3}\times N_{\rm run}$) in
unit of MeV/$c^{2}$, where $N_{\rm run}$ is the run number, and the
largest value from error propagation is taken as the corresponding
statistical uncertainty. For other data samples,
$M^{\rm obs}(\mu^{+}\mu^{-})$ remains stable, and the average value is
used to calculate $E_{\rm cms}$. The samples 4009$^{1}$
(4420$^{2}$) and 4009$^{2}$ (4420$^{3}$) are separated because they show
a sudden drop in the average energies. Table~\ref{MDimu_list} (column
5) summarizes the measured $M^{\rm obs}(\mu^{+}\mu^{-})$ for each sample.

The $E_{\rm cms}$ is finally obtained by adding the energy-dependent
mass shift $\overline{\Delta M}_{\rm ISR/FSR}$ due to ISR/FSR obtained
in Section~\ref{correct_dM} to the measured
$M^{\rm obs}(\mu^{+}\mu^{-})$. The measured $E_{\rm cms}$ is listed in
Table~\ref{MDimu_list} (column 6);  the systematic uncertainty
will be discussed in Section~\ref{Sys_err}.

\begin{table}[t]
  %\footnotesize
  \centering
  \caption{\label{MDimu_result}Weighted average $E_{\rm cms}$ for
    all data samples. The first uncertainty is statistical, and the
    second is systematic.}
  \begin{tabular}{ccc}
\hline
  ID && Weighted average $E_{\rm cms}$ (MeV)\\
\hline
  3810 &&3807.65$\pm$0.10$\pm$0.58   \\
  3900 &&3896.24$\pm$0.11$\pm$0.72   \\
  4009 &&4007.62$\pm$0.05$\pm$0.66   \\
  4090 &&4085.45$\pm$0.14$\pm$0.66   \\
  4190 &&4188.59$\pm$0.15$\pm$0.68   \\
  4210 &&4207.73$\pm$0.14$\pm$0.61   \\
  4220 &&4217.13$\pm$0.14$\pm$0.67    \\
  4230 &&4226.26$\pm$0.04$\pm$0.65   \\
  4245 &&4241.66$\pm$0.12$\pm$0.73  \\
  4260 &&4257.97$\pm$0.04$\pm$0.66  \\
  4310 &&4307.89$\pm$0.17$\pm$0.63  \\
  4360 &&4358.26$\pm$0.05$\pm$0.62   \\
  4390 &&4387.40$\pm$0.17$\pm$0.65   \\
  4420 &&4415.58$\pm$0.04$\pm$0.72   \\
  4470 &&4467.06$\pm$0.11$\pm$0.73  \\
  4530 &&4527.14$\pm$0.11$\pm$0.72   \\
  4575 &&4574.50$\pm$0.18$\pm$0.70   \\
  4600 &&4599.53$\pm$0.07$\pm$0.74  \\
\hline

\end{tabular}
\end{table}

Each of the data sets 4009, 4230, 4260, and 4420 is split into several
sub-samples. We calculate the luminosity-weighted average $E_{\rm cms}$
for each, and the largest systematic uncertainty of the samples is
taken as the systematic uncertainty. In Table~\ref{MDimu_result},
we summarize the weighted average $E_{\rm cms}$ for all data samples.

\section{\boldmath cross check}

The processes of $e^{+}e^{-}\rightarrow\pi^{+}\pi^{-}K^{+}K^{-}$ and
$e^{+}e^{-}\rightarrow\pi^{+}\pi^{-}p\bar{p}$ are used to check the
measurement of the $E_{\rm cms}$. Similar to the di-muon process
$e^{+}e^{-}\rightarrow\gamma_{\rm ISR/FSR}\mu^{+}\mu^{-}$, the
$E_{\rm cms}$ of the initial $e^{+}e^{-}$ system is estimated by the
corrected invariant masses of the final state
particles $M^{\rm cor}(\pi^{+}\pi^{-}K^{+}K^{-})$ and
$M^{\rm cor}(\pi^{+}\pi^{-}p\bar{p})$. The measurement of the low momentum
charged tracks is validated using the decay channels
$D^{0}\rightarrow K^{-}\pi^{+}$ and
$\bar{D}^{0}\rightarrow K^{+}\pi^{-}$. The measured mass,
$M^{\rm obs}(K^{-}\pi^{+}/K^{+}\pi^{-}) = 1864.00\pm0.7$ MeV
(statistical uncertainty only) is consistent with the nominal
$D^{0}/\bar{D}^{0}$ mass~\cite{PDG} with a deviation of $0.84\pm0.71$
MeV. Both the corrected $M^{\rm cor}(\pi^{+}\pi^{-}K^{+}K^{-})$ and
$M^{\rm cor}(\pi^{+}\pi^{-}p\bar{p})$ are found to be consistent with
$E_{\rm cms}$ obtained using the di-muon process, with the largest
deviation of $0.53\pm0.75$ MeV found in sample 4420.

\section{\boldmath Systematic uncertainties}\label{Sys_err}

The systematic uncertainty in $E_{\rm cms}$ in this analysis is
estimated by considering the
uncertainties from the momentum measurement of the $\mu^{\pm}$, the
estimation of the mass shift $\Delta M_{\rm ISR/FSR}$ due to ISR/FSR,
the generator, and the corresponding fit procedure.

We use the $J/\psi$ invariant mass via the process
$J/\psi\rightarrow\mu^{+}\mu^{-}$ to check the momentum
reconstruction. The measured $J/\psi$ mass corrected for FSR effects at
each energy, $M^{\rm cor}(J/\psi)$, is close to the nominal $J/\psi$
mass. To be conservative, we use a first-order polynomial to fit the
$M^{\rm cor}(J/\psi)$ versus $E_{\rm cms}$ distribution, and find the
largest difference in the $J/\psi$ mass between the fit result and
the nominal value to be 0.34 MeV/$c^{2}$. We take
$\frac{0.34}{3096.92}=0.011\%$ as the systematic uncertainty due to
the momentum measurement.

The mass shift $\Delta M_{\rm ISR/FSR}$ due to ISR/FSR is
$E_{\rm cms}$ dependent, and is obtained from MC samples with
50,000 generated events each.   The standard deviation of the
distribution of $\Delta M_{\rm ISR/FSR}$ versus $E_{\rm cms}$ is given
by
\begin{equation}\label{standard_error}
  \sigma = \sqrt{\frac{\Sigma(\Delta M_{\rm ISR/FSR}-\overline{\Delta M}_{\rm ISR/FSR})^{2}}{N-1}} = 0.37 ~{\rm MeV}/c^{2},
\end{equation}
where $\overline{\Delta M}_{\rm ISR/FSR}$ is the value from the fit
(Fig.~\ref{dDimu_all_MC}), and $N$ is the number of points in
Fig.~\ref{dDimu_all_MC}.  A value of 0.37 MeV/$c^{2}$ is taken as
systematic uncertainty due to the ISR/FSR correction.

Additionally, we use different generator versions ({\sc babayaga3.5}
and {\sc babayaga@NLO}) to estimate the mass shift
$\Delta M_{\rm ISR/FSR}$.  The averaged difference in
$\Delta M_{\rm ISR/FSR}$ from the two generators is $0.036\pm0.067$
MeV/$c^{2}$, which reflects the contribution to the systematic
uncertainty of the ISR/FSR correction from the generator;  it is
negligibly small.

The $M^{\rm obs}(\mu^{+}\mu^{-})$ is measured run-by-run and is found
to be stable
during data-taking for most samples. For the runs in each sample
(except for the samples of $4230^{2}$ and $4260^{1}$, which are described by
a first-order polynomial), the average $E_{\rm cms}$ is provided to
reduce the statistical fluctuation. If the energy shifts gradually
during the data-taking, the simple average value will cause a systematic
uncertainty. To estimate this systematic error for each sample, we fit
the distribution of $M^{\rm obs}(\mu^{+}\mu^{-})$ versus run-number by a
first-order polynomial and take the largest difference between the
fitting result and the average value, less than 0.25 MeV on average,
as the systematic uncertainty.

The uncertainties from other sources, such as background and event
selection, are negligible. Assuming all the sources of systematic
uncertainty are independent, the total systematic uncertainty is
obtained by adding all items in quadrature, which is listed in
Table~\ref{MDimu_list} (column 6). The uncertainty is smaller than 0.8
MeV for all the data samples.

\section{\boldmath Summary}

The center-of-mass energies of the data taken from 2011 to 2014 for
the studies of the charmonium-like and higher excited charmonium
states are measured with the di-muon process
$e^{+}e^{-}\rightarrow\gamma_{\rm ISR/FSR}\mu^{+}\mu^{-}$.  The
corresponding statistical uncertainty is very small, and the
systematic uncertainty is found to be less than 0.8 MeV. The measured
$E_{\rm cms}$ is validated by the processes
$e^{+}e^{-}\rightarrow\pi^{+}\pi^{-}K^{+}K^{-}$ and
$e^{+}e^{-}\rightarrow\pi^{+}\pi^{-}p\bar{p}$. The stability of
$E_{\rm cms}$ over time for the data samples is also examined. For the
samples 4009, 4230, 4260, 4420, we also give the luminosity-weighted
average $E_{\rm cms}$. The results are essential for the discovery of
new states and investigation of the transition of charmonium and
charmonium-like states~\cite{Z3900,Z4020,Z3885,Z4025}.

\acknowledgements

The BESIII collaboration thanks the staff of BEPCII and the IHEP computing center for their strong support. This work is supported in part by National Key Basic Research Program of China under Contract No. 2015CB856700; National Natural Science Foundation of China (NSFC) under Contracts Nos. 11125525, 11235011, 11322544, 11335008, 11425524; the Chinese Academy of Sciences (CAS) Large-Scale Scientific Facility Program; the CAS Center for Excellence in Particle Physics (CCEPP); the Collaborative Innovation Center for Particles and Interactions (CICPI); Joint Large-Scale Scientific Facility Funds of the NSFC and CAS under Contracts Nos. 11179007, U1232201, U1332201; CAS under Contracts Nos. KJCX2-YW-N29, KJCX2-YW-N45; 100 Talents Program of CAS; National 1000 Talents Program of China; INPAC and Shanghai Key Laboratory for Particle Physics and Cosmology; German Research Foundation DFG under Contract No. Collaborative Research Center CRC-1044; Istituto Nazionale di Fisica Nucleare, Italy; Ministry of Development of Turkey under Contract No. DPT2006K-120470; Russian Foundation for Basic Research under Contract No. 14-07-91152; The Swedish Research Council; U. S. Department of Energy under Contracts Nos. DE-FG02-04ER41291, DE-FG02-05ER41374, DE-FG02-94ER40823, DESC0010118; U.S. National Science Foundation; University of Groningen (RuG) and the Helmholtzzentrum fuer Schwerionenforschung GmbH (GSI), Darmstadt; WCU Program of National Research Foundation of Korea under Contract No. R32-2008-000-10155-0.

\end{CJK*}
\end{document}

%% file: author_20150731.tex
\author{
  \begin{small}
    \begin{center}
      M.~Ablikim$^{1}$, M.~N.~Achasov$^{9,f}$, X.~C.~Ai$^{1}$,
      O.~Albayrak$^{5}$, M.~Albrecht$^{4}$, D.~J.~Ambrose$^{44}$,
      A.~Amoroso$^{49A,49C}$, F.~F.~An$^{1}$, Q.~An$^{46,a}$,
      J.~Z.~Bai$^{1}$
% ...
\\
    \vspace{0.2cm}
    (BESIII Collaboration)\\
      \vspace{0.2cm} {\it
        $^{1}$ Institute of High Energy Physics, Beijing 100049, People's Republic of China\\
% ...
$^{4}$ Bochum Ruhr-University, D-44780 Bochum, Germany\\
          $^{5}$ Carnegie Mellon University, Pittsburgh, Pennsylvania 15213, USA\\
% ...
$^{9}$ G.I. Budker Institute of Nuclear Physics SB RAS (BINP), Novosibirsk 630090, Russia\\
% ...
$^{44}$ University of Rochester, Rochester, New York 14627, USA\\
% ...
$^{46}$ University of Science and Technology of China, Hefei 230026, People's Republic of China\\
% ...
$^{49}$ (A)University of Turin, I-10125, Turin, Italy; (B)University of Eastern Piedmont, I-15121, Alessandria, Italy; (C)INFN, I-10125, Turin, Italy\\
% ...
$^{a}$ Also at State Key Laboratory of Particle Detection and Electronics, Beijing 100049, Hefei 230026, People's Republic of China\\
% ...
$^{f}$ Also at the Novosibirsk State University, Novosibirsk, 630090, Russia\\
% ...
      }\end{center}
\vspace{0.6cm}
\end{small}
}
%\affiliation{}